\documentclass[fleqn,usenatbib,useAMS]{mnras}

\usepackage{graphicx}	
\usepackage{amsmath}	
\usepackage{multicol}        
\usepackage{bm}		
\usepackage{pdflscape}	
\usepackage{fix-cm}
\usepackage{float}
\usepackage{booktabs}
\usepackage{caption}

\newcommand{\rhomean}{\langle\rho\rangle}

\newcommand{\cs}{c_\mathrm{s}}
\newcommand{\ekin}{E_{\mathrm{kin}}}
\newcommand{\epskin}{\varepsilon_{\mathrm{kin}}}
\newcommand{\strain}{\varepsilon_{\bm{\mathcal{S}}}}
\newcommand{\epsinj}{\varepsilon_{\mathrm{inj}}}
\newcommand{\mach}{{\mathcal{M}}}
\newcommand{\vort}{\vert\nabla\times \mathbf{v}\vert}

\newcommand{\del}{\partial}
\newcommand{\dt}{\Delta t}
\newcommand{\dx}{\Delta x}
\newcommand{\Reynolds}{\mathrm{Re}}
\newcommand{\tturb}{t_{\mathrm{turb}}}
\newcommand{\vturb}{v_{\mathrm{turb}}}
\newcommand{\kdriv}{k_{\mathrm{driv}}}
\newcommand{\ldriv}{\ell_{\mathrm{driv}}}
\newcommand{\injdissdt}{\Delta t_\mathrm{inj-diss}}
\newcommand{\injdissC}{\mathcal{C}_\mathrm{inj-diss}}
\newcommand{\abs}[1]{\vert{#1}\vert}
\newcommand{\emdash}{\,---\,}
\newcommand{\inquotes}[1]{`{#1}'}

\usepackage[T1]{fontenc}
\usepackage{ae,aecompl}

\usepackage{newtxtext,newtxmath}

\usepackage{scalerel,tikz}
\usetikzlibrary{svg.path}
\definecolor{orcidlogocol}{HTML}{A6CE39}
\tikzset{orcidlogo/.pic={
 \fill[orcidlogocol] svg{M256,128c0,70.7-57.3,128-128,128C57.3,256,0,198.7,0,128C0,57.3,57.3,0,128,0C198.7,0,256,57.3,256,128z};
 \fill[white] svg{M86.3,186.2H70.9V79.1h15.4v48.4V186.2z}
 svg{M108.9,79.1h41.6c39.6,0,57,28.3,57,53.6c0,27.5-21.5,53.6-56.8,53.6h-41.8V79.1z M124.3,172.4h24.5c34.9,0,42.9-26.5,42.9-39.7c0-21.5-13.7-39.7-43.7-39.7h-23.7V172.4z}
 svg{M88.7,56.8c0,5.5-4.5,10.1-10.1,10.1c-5.6,0-10.1-4.6-10.1-10.1c0-5.6,4.5-10.1,10.1-10.1C84.2,46.7,88.7,51.3,88.7,56.8z};
}}
\newcommand\orcidicon[1]{\href{https://orcid.org/#1}{\mbox{\scalerel*{
\begin{tikzpicture}[yscale=-1,transform shape]
\pic{orcidlogo};
\end{tikzpicture}
}{|}}}}

\title[Dissipative turbulent structures]{The statistics and structure of dissipation in subsonic and supersonic turbulence}

\author[Troccoli \& Federrath]{
Edward Troccoli$^{\orcidicon{0009-0004-3213-2216}\,1}$\thanks{Email: \href{mailto:Edward.Troccoli@anu.edu.au}{edward.troccoli@anu.edu.au}} \&
Christoph Federrath$^{\orcidicon{0000-0002-0706-2306}\,1}$\thanks{Email: \href{mailto:christoph.federrath@anu.edu.au}{christoph.federrath@anu.edu.au}}
\\
$^{1}$Research School of Astronomy and Astrophysics, Australian National University, Canberra, ACT 2611, Australia
}

\date{\today}

\pubyear{2025}

\begin{document}
\label{firstpage}
\pagerange{\pageref{firstpage}--\pageref{lastpage}}
\maketitle
    
    \begin{abstract}
    Turbulence plays a critical role in the atmosphere, oceans, engineering, and astrophysics. The dissipation (heating) induced by turbulent flows is particularly important for the thermodynamics and chemistry of interstellar clouds, yet its structure and statistics remain poorly understood. Using high-resolution turbulence simulations with controlled explicit viscosity, we study the kinetic energy dissipation rate, $\epskin$, across subsonic and supersonic regimes. We find that dissipation lags large-scale kinetic energy injection events by $1.64\pm0.21$ and $0.48\pm0.07$ turbulent turnover times in subsonic and supersonic turbulence, respectively. Correlations show $\epskin\propto\vort^2$ (vorticity squared) in the subsonic regime, where density fluctuations are negligible, while in the supersonic regime dissipation is primarily correlated with density, $\epskin\propto\rho^{3/2}$. A spectral analysis demonstrates that achieving numerical convergence of $\epskin$ across all scales is challenging, especially in the subsonic case, even at $2048^3$ resolution. Nonetheless, subsonic dissipation is clearly localised on small vorticity-dominated scales, while supersonic dissipation spans many scales, combining elongated, thin shocks with small-scale vorticity. Finally, we determine the fractal dimension of $\epskin$. In the subsonic regime, intense dissipation is predominantly organised in flattened vortex filaments embedded in thin shearing layers on small scales, becoming more volume-filling at larger scales. In the supersonic regime, $\epskin$ exhibits a fractal dimension between 1 and 2 across nearly all scales, likely reflecting shock surfaces and their intersections forming filaments.
    \end{abstract}



\section{Introduction}

Turbulence remains a largely unsolved physical process \citep{LandauLifchitz1959,DavidsonKanedaMoffattSreenivasan2011}. It is fundamentally chaotic, characterised by an energy cascade from large scales (driving) to small scales (dissipation). The study of turbulence is relevant in both industrial and academic contexts, with key applications in accurate weather prediction, climate modelling, ocean mixing \citep{Gargett1989,SohailEtAl2019}, and in engines and turbines \citep{GiustiMastorakos2019,LeggettZhaoSandberg2023}. Moreover, many astrophysical phenomena are strongly influenced, or even controlled, by turbulent flows, such as those in the interstellar medium (ISM) \citep{Ferriere2001,ScaloElmegreen2004}, where turbulence plays a key role in star formation \citep{ElmegreenScalo2004,MacLowKlessen2004,McKeeOstriker2007,HennebelleFalgarone2012,PadoanEtAl2014,Federrath2018,MathewFederrathSeta2023}. Despite its importance, turbulence{\emdash}and in particular its dissipation{\emdash}remains poorly understood.

The dissipation of kinetic energy has significant implications \citep{UritskyPouquetRosenberg2010,RichardLesaffreFalgaroneLehmann2022} in astrophysics. In general, turbulent dissipation governs the rate at which large-scale injected energy is converted into heat, thereby catalysing chemical reactions and altering the thermodynamics of star-forming clouds \citep{HilyBlantFalgaronePety2008,FalgaronePetyHilyBlant2009}. Dissipation is a highly intermittent process \citep{PorterPouquetWoodward2002,MomferratosLesaffreFalgaronePineaudesForêts2014,Zhdankin2014}, observed primarily on small scales \citep{Kolmogorov1962,FalgaronePhillips1990,MeneveauSreenivasan1991,FalgaronePugetPerault1992},
a feature that has been confirmed in laboratory experiments
\citep{AnselmetAntoniaDanaila2001}.It occurs via two main pathways: small-scale vortices \citep{DouadyCouderBrachet1991,Frisch1995}, and{\emdash}in the case of supersonic turbulence relevant for molecular clouds{\emdash}via shocks \citep{Burgers1948,LehmannFederrathWardle2016,FederrathEtAl2021}. These structures thus provide signatures of dissipation and are a main focus of this study.

Small-scale vortices associated with dissipation are typically found in subsonic turbulence and tend to occupy 3D space in a fractal manner \citep{MeneveauSreenivasan1991}. Specifically, the observed correlation with vorticity arises because most of the dissipation occurs in the shear layers wrapped around vortex tubes, where the turbulent rate of strain is strongest \citep{MoffattKidaOhkitani1994}, producing thin, ribbon-like dissipative structures \citep{MoisyJimenez2004}. In contrast, shocks dominate supersonic turbulence, appearing as elongated, thin sheets which, upon collision, form filamentary structures at their intersections. Shock collisions therefore localise dissipation into intermittent structures \citep{PorterPouquetWoodward2002} and introduce an additional channel for supersonic dissipation beyond vortex-driven processes.

In this work, we determine the structure and statistics of dissipation by implementing a new method to measure the rate at which local kinetic energy dissipates in fully-developed turbulence. We use numerical simulations on discretised grids up to $2048^3$~cells, evolving the kinetic energy equation of hydrodynamics. Combined with the standard hydrodynamical equations, particularly the momentum equation, this allows us to compute the local dissipation rate. At each time step, we calculate the kinetic energy conventionally using the momentum and continuity equations, and compare it to the evolved kinetic energy equation. The difference between these quantities, per unit time, directly yields the dissipation rate. This formulation naturally accounts for both explicit (viscous) and numerical (resolution-dependent) dissipation. To minimise the latter, we include explicit viscosity terms and perform simulations at varying grid resolutions to assess convergence.

The paper is organised as follows. In Section~\ref{sec:methods}, we introduce the turbulence simulation techniques employed. Section~\ref{sec:results} presents the main results, including visualisations of simulated quantities, their time evolution, correlations of the dissipation rate, power spectra of dissipation and kinetic energy, and a fractal dimension analysis of dissipation. Section~\ref{sec:conclusions} provides a summary and outlines future directions. Derivations of useful equations and supplementary figures are given in Appendix~\ref{sec:derivation} and~\ref{sec:l2norm}, respectively.

\section{Methods} \label{sec:methods}

\subsection{Hydrodynamic equations} \label{sec:hd}

The turbulent systems studied here are described by the compressible, three-dimensional, hydrodynamic (HD) equations, given by
\begin{align}
    \del_t\rho+\nabla\cdot(\rho\mathbf{v})&=0 \label{eq:masscons},\\
    \del_t(\rho\mathbf{v})+\nabla\cdot(\rho\mathbf{v}\otimes\mathbf{v})+\nabla p&=\nabla\cdot(2\nu\rho\bm{\mathcal{S}})+\rho\mathbf{F} \label{eq:momentum},\\
    \del_tE+\nabla[(E+p)\mathbf{v}]&=\nabla\cdot[2\nu\rho\mathbf{v}\cdot\bm{\mathcal{S}}] + \rho\mathbf{v}\cdot\mathbf{F}\label{eq:fullenergy},
\end{align}
where $\rho$ denotes the density, $\mathbf{v}$ the velocity, $p$ the thermal pressure, and $\otimes$ denotes the outer product of two vectors, i.e., $\mathbf{q}\otimes\mathbf{q}=\mathbf{q}\mathbf{q}^{\mathrm{T}}$. The total energy density, $E=\rho e_{\textrm{int}}+\rho\vert\mathbf{v}\vert^2/2$, is the sum of the internal and kinetic energy densities, respectively. Explicit viscous dissipation is included through the traceless rate of strain tensor, $\mathbf{\mathcal{S}}_{ij}=(1/2)(\del_{i}v_{j}+\del_{j}v_{i})-(1/3)\delta_{ij}\nabla\cdot\mathbf{v}$, with the amount of kinematic viscosity determined by $\nu$. To close the HD equations we use a polytropic equation of state (EOS) $p=\cs^2\rho$, such that the gas remains isothermal with a constant sound speed $\cs$. It should be noted that this EOS implies that the baroclinic vorticity is zero \citep{MeeBrandenburg2006}. As turbulence transports energy from large to small scales, where it is dissipated, we need to drive the turbulence on large scales, which is achieved by the acceleration field described through the $\rho\mathbf{F}$ term in Equation \eqref{eq:momentum}. Here, we work in the purely HD case (no magnetic fields), and leave the case including magnetic fields and its dissipation for future work.

\subsection{Kinetic energy equation}\label{sec:ekin}

Since we are primarily interested in studying the rate of kinetic energy dissipation, we first need to consider the evolution equation of the kinetic energy, $\ekin$. The evolution equation for $\ekin$ is derived by first computing the dot product of the momentum Eq.~(\ref{eq:momentum}) with $\mathbf{v}$. After a few steps involving use of the continuity Eq.~(\ref{eq:masscons}), we find (see Appendix~\ref{sec:derivation} for details),
\begin{align}
    \del_t\ekin+\nabla\cdot\left[(\ekin+p)\mathbf{v}\right] = p\nabla\cdot\mathbf{v} \label{eq:ekin}.
\end{align}
This equation describes the evolution of kinetic energy without any dissipation, neither from physical, explicit terms, through the viscosity set in Eqs.~(\ref{eq:momentum}) and~(\ref{eq:fullenergy}), nor by numerical effects. We also note that the turbulence forcing term ($\mathbf{F}$) is not included here, because we want to measure dissipation, not injection. In fact, the forcing term is operator split from the HD time step, and as explained below, it is therefore not part of the HD update and consequently not relevant for computing the dissipation rate.

\subsection{Dissipation rate} \label{sec:epskin}

Using Eq.~(\ref{eq:ekin}), we obtain the \inquotes{perfectly} evolved kinetic energy, without contributions from dissipation, i.e., from viscous numerical and explicit terms, which act to reduce $\ekin$ via transfer to thermal energy on small scales. However, in a standard hydrodynamic code, Eq.~(\ref{eq:ekin}) is normally not evolved. Instead, the kinetic energy is obtained after solving the continuity and momentum Eqs.~(\ref{eq:masscons}) and~(\ref{eq:momentum}) as $(1/2)\rho\abs{\mathbf{v}}^2$, and this quantity is subject to dissipation. Thus, by evolving the standard set of hydro Eqs.~(\ref{eq:masscons})--(\ref{eq:fullenergy}) together with Eq.~(\ref{eq:ekin}), we obtain the dissipated kinetic energy as the difference between $\ekin$ from Eq.~(\ref{eq:ekin}) and $(1/2)\rho\abs{\mathbf{v}}^2$.

In practice, at the start of a HD time step $\dt$, at time $t$, we (re)set $\ekin(t)=(1/2)\rho(t)\abs{\mathbf{v}(t)}^2$ and update it via Eq.~(\ref{eq:ekin}) to obtain $\ekin(t+\dt)$. Meanwhile, the standard set of HD equations is solved as well, in particular Eqs.~(\ref{eq:masscons}) and~(\ref{eq:momentum}), to obtain the updated density, $\rho(t+\dt)$ and velocity, $\mathbf{v}(t+\dt)$, which defines a kinetic energy after the HD time step $(1/2)\,\rho(t+\dt)\,\abs{\mathbf{v}(t+\dt)}^2$, but this one was subject to dissipation. The difference between these two quantities then defines the kinetic energy dissipation rate $\epskin$ through
\begin{equation}
    \epskin(t+\dt)\equiv\frac{\ekin(t+\dt)-(1/2)\,\rho(t+\dt)\,\abs{\mathbf{v}(t+\dt)}^2}{\dt}.
\end{equation}
The quantity $\epskin$ is therefore defined locally in each grid cell. We can obtain the total dissipation rate per time step by integrating over space.

\subsection{Numerical simulations}

Here we summarise the main numerical methods used for the turbulence simulations. Tab.~\ref{tab:sim_params} provides a list of all the simulations performed in this work and their main parameters. Additionally, selected measurements are shown to aid our resolution studies.

\begin{table*}
    \centering
    \caption{Main simulation parameters (columns~1--4) and measurements (columns~5--7).}
    \label{tab:sim_params}
    \renewcommand{\arraystretch}{1.0} 
    \setlength{\tabcolsep}{5.0pt} 
    \begin{tabular}{lcccccc}
        \hline
        Model & Mach number $(\mach)$ & Grid resolution $(N^3)$ & Viscosity $(\nu)$ & Dissipation rate $(\epskin)$ & Inj.-diss.~lag $(\injdissdt)$ & Small-scale fractal \\
        &  &  & $[\ldriv^2\tturb^{-1}]$ & $[\rhomean\mach^2\cs^2\tturb^{-1}]$ & $[\tturb]$ & dimension ($D$) \\
        (1) & (2) & (3) & (4) & (5) & (6) & (7) \\
        \hline
        Sub2048 & 0.2 & $2048^3$ & $4\times10^{-5}$ & $0.31\pm0.05$ & $1.64\pm0.21$  & $1.99\pm0.07$ \\
        Sub1024 & 0.2 & $1024^3$ & $4\times10^{-5}$ & $0.35\pm0.06$ & $1.66\pm0.22$ & $1.87\pm0.08 $ \\
        Sub512  & 0.2 &  $512^3$ & $4\times10^{-5}$ & $0.53\pm0.08$ & $1.79\pm0.20$ & $1.46\pm0.07$ \\
        Sub256  & 0.2 &  $256^3$ & $4\times10^{-5}$ & $0.89\pm0.12$ & $1.83\pm0.16$ & $1.16\pm0.07$ \\
        \hline
        Sup2048   & 5 & $2048^3$ & $1\times10^{-3}$ & $0.56\pm0.09$ & $0.48\pm0.07$ & $1.60\pm0.07$ \\
        Sup1024   & 5 & $1024^3$ & $1\times10^{-3}$ & $0.54\pm0.08$ & $0.49\pm0.09$ & $1.56\pm0.06$ \\
        Sup512    & 5 &  $512^3$ & $1\times10^{-3}$ & $0.52\pm0.07$ & $0.41\pm0.01$ & $1.46\pm0.05$ \\
        Sup256    & 5 &  $256^3$ & $1\times10^{-3}$ & $0.48\pm0.06$ & $0.41\pm0.02$ & $1.01\pm0.02$ \\
        \hline
    \end{tabular}
\end{table*}

\subsubsection{Basic simulation code}

We solve the HD equations (Eqs.~\ref{eq:masscons}--\ref{eq:fullenergy}) with a modified version of the astrophysical code FLASH \citep{fryxell_flash_2000} (based on version~4), using the HLL5R 5-wave approximate Riemann solver \citep{bouchut_multiwave_2010,WaaganFederrathKlingenberg2011}. This is done on a uniformly discretised, triply-periodic, 3D grid with side length $L$ and $N^3$~grid cells.

\subsubsection{Turbulence driving} \label{sec:turbdriv}

To drive turbulence and maintain a constant rms kinetic energy injection rate $\epsinj$, we require an acceleration field, $\mathbf{F}$, which appears in Eq.~\eqref{eq:momentum}. To construct $\mathbf{F}$ we use an Ornstein-Uhlenbeck (OU) process \citep{EswaranPope1988,SchmidtEtAl2006b,FederrathDuvalKlessenSchmidtMacLow2010} as implemented in the publicly available turbulence driving module {\sc TurbGen} \citep{FederrathEtAl2022ascl}, which generates a time sequence of random Fourier modes via the OU process. All simulations presented in this paper employ {\sc TurbGen} with a turbulence driving pattern with characteristic driving wave number $\kdriv=2$, where $k$ is in units of $2\pi/L$, i.e., the driving scale is $\ldriv=1/2$ in units of $L$. Thus, energy is primarily injected on half the box scale. The driving amplitude falls off parabolically on both sides of the peak, reaching exactly zero at $k=1$ and $3$. Thus, only a narrow range of wave numbers is driven, maximising the self-consistent development of fully-developed turbulence for $k\gtrsim3$.

The large-scale eddy turnover time is the characteristic time scale of the turbulence, defined as $\tturb=\ldriv/\vturb$, where $\vturb$ is the turbulent velocity dispersion on the driving scale. For brevity, $\tturb$ will be referred to as the turnover time throughout the paper. The driving amplitude is set such that the sonic Mach number, $\mach=\vturb/\cs=0.2$ and $5$ for the subsonic and supersonic simulations, respectively. Turbulence becomes fully developed after $\sim2\tturb$ \citep{Federrath2010,PriceFederrath2010}, and therefore all measurements of turbulent quantities will focus on $t/\tturb\geq2$.

\subsubsection{Initial conditions and simulation parameters}

We compare subsonic ($\mach=0.2$) and supersonic ($\mach=5$) simulations, and study numerical convergence by using grid resolutions of $N=256$, $512$, $1024$, and $2048$, as summarised in Tab.~(\ref{tab:sim_params}). We set the sound speed $\cs=1$ along with the initial density $\rho=\rhomean=1$. Therefore, all quantities with units of length, time, velocity, density, and wave number are presented in units of $\ldriv$, $\tturb$, $\cs$, $\rhomean$ and $2\pi/L$, respectively. Kinetic energy is expressed in units of $\rhomean\mach^2\cs^2$, and energy dissipation/injection rates in units of $\rhomean\mach^2\cs^2\tturb^{-1}$. Spatial derivatives are in units of $\dx^{-1}$, with the cell side length $\dx=L/N$. 

Across all simulations, we prescribe a constant kinematic viscosity $\nu$, \citep{LesaffreEtAl2020,KrielEtAl2022} as opposed to a constant dynamic viscosity, $\mu=\nu\rho$, so that we can set and control a global Reynolds number
\begin{equation}
\Reynolds = \frac{\ldriv\vturb}{\nu}.
\end{equation}
This allows us to set a fixed $\Reynolds$ in both the subsonic and supersonic set of simulations. Using a fixed dynamic viscosity would not allow us to set a clean Reynolds number, as $\rho$ varies substantially in the supersonic case. Setting a fixed $\nu$ instead of a fixed $\mu$ further ensures that the dissipation wave number, $k_{\nu}$, is a global quantity, determined by the Reynolds number \citep{KrielEtAl2025}. Moreover, there is an additional complication with setting $\mu$ constant in supersonic turbulence, because with constant $\mu$, the time step $\dt\sim\dx^2/\nu=\dx^2\rho/\mu$ would become density-dependent. This leads to the computations becoming prohibitively expensive given the very large density variations present in supersonic turbulence, with under-densities of several orders of magnitude relative to the mean density.

\section{Results} \label{sec:results}

\subsection{Structure and morphology of fluid quantities} \label{sec:movie}

\begin{figure*}
\centering
\def\arraystretch{0.0}
\setlength{\tabcolsep}{0pt}
 \includegraphics[width=1\textwidth]{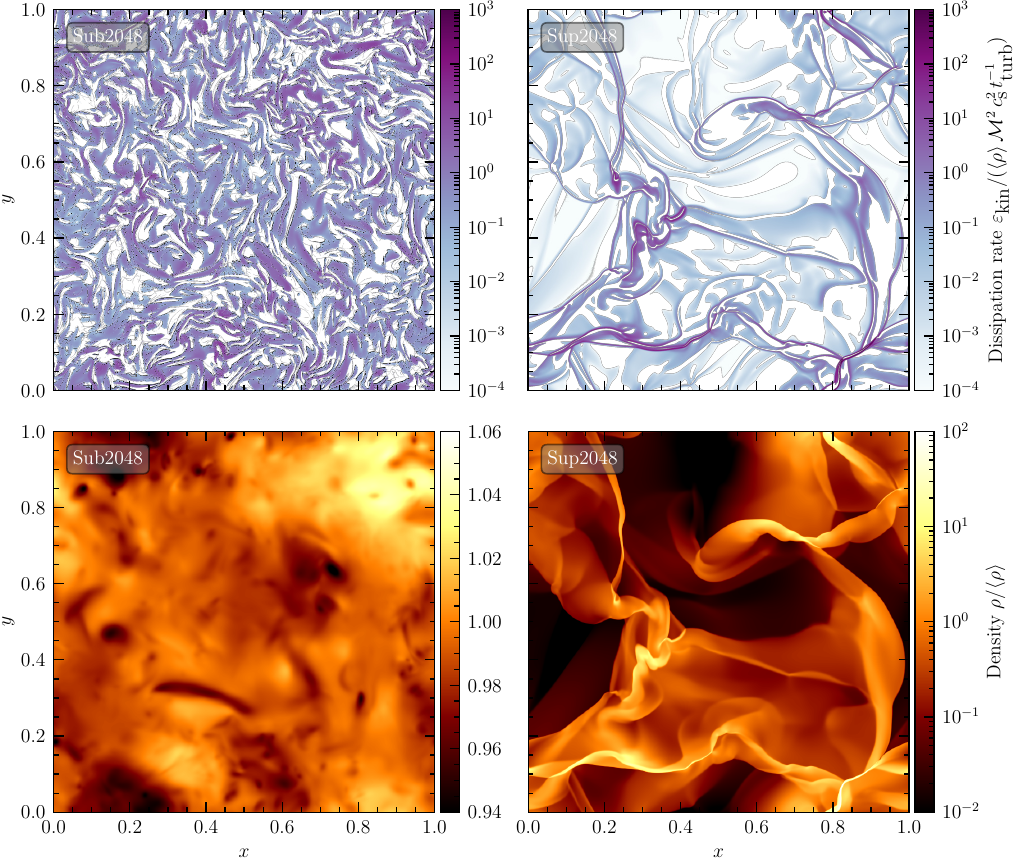} 
\caption{Slices through the kinetic energy dissipation rate $\epskin$ (top panels) and the gas density (bottom panels), in the subsonic regime ($\mach=0.2$; left-hand panels) and the supersonic regime ($\mach=5$; right-hand panels), at $t=5\,\tturb$, i.e., in the fully-developed state of turbulence, for the simulations with $2048^3$~grid cells (cf.~Tab.~\ref{tab:sim_params}). We see that for subsonic turbulence, $\epskin$ is more volume filling than in supersonic turbulence, and the dissipation is organised into relatively thick elongated structures. In contrast, the supersonic case shows very thin high-dissipation structures, likely associated with shocks, as reflected in the density field.}
\label{fig:2x2movieframes}
\end{figure*}

We first seek to gain a qualitative understanding of the kinetic energy dissipation rate $\epskin$, which we defined in Section~\ref{sec:epskin}. This is presented in Figure~(\ref{fig:2x2movieframes}), which shows $\epskin$ (top panels) for a slice in the $xy$-plane, alongside the density (bottom panels) for the corresponding slice in both the subsonic (left) and supersonic (right) simulations with $2048^3$~cells.

We see that the dissipation rate in the subsonic regime is visually volume filling compared to very localised shock filaments in supersonic flows. Nevertheless, as a function of space, there are orders of magnitude variation in $\epskin$ in both regimes. In the subsonic case, dissipation structures are present as thick, somewhat elongated structures, and are frequent throughout the domain. This is in contrast to the supersonic case where the dissipative structures span the domain as very long and narrow shocks. It is important to note that as we produce slices of the $xy$ plane, then what appear to be line filaments in the supersonic dissipation in Figure~(\ref{fig:2x2movieframes}) may not all be actual lines (filaments), but can also be slices through shock surfaces.

We also observe that the bottom left panel of Figure~(\ref{fig:2x2movieframes}) shows negligible variation in the density, as expected for subsonic turbulence \citep{KonstandinEtAl2012ApJ}. Therefore, we do not expect a correlation of the dissipation rate with density, but instead perhaps with the vorticity of the gas \citep{Frisch1995}. In contrast, the supersonic case exhibits large density variations with dense shocks seemingly in correspondence with the the high-$\epskin$ structures. Moreover, throughout the domain, we observe a complex network of interacting shocks, which intermittently collide, forming knot-like structures. Within these collisions, the gas behaves highly chaotically and introduces vorticity at the site of the collision due to the contribution of a logarithmic density gradient term in the vorticity evolution equation \citep{MeeBrandenburg2006,Federrath2011}. This indicates that these collisions may provide an additional channel for dissipation.

In the following, we provide a quantitative investigation of the  correlations of $\epskin$ with density and vorticity, as well as the dimensionality of these structures through fractal analysis. Before that, however, we examine the time evolution and temporal correlations between the injection and dissipation of kinetic energy.

\subsection{Time evolution} \label{sec:timevol}

\begin{figure*}
\centering
\setlength{\tabcolsep}{1pt}
\renewcommand{\arraystretch}{1}
	\includegraphics[width=1\textwidth]{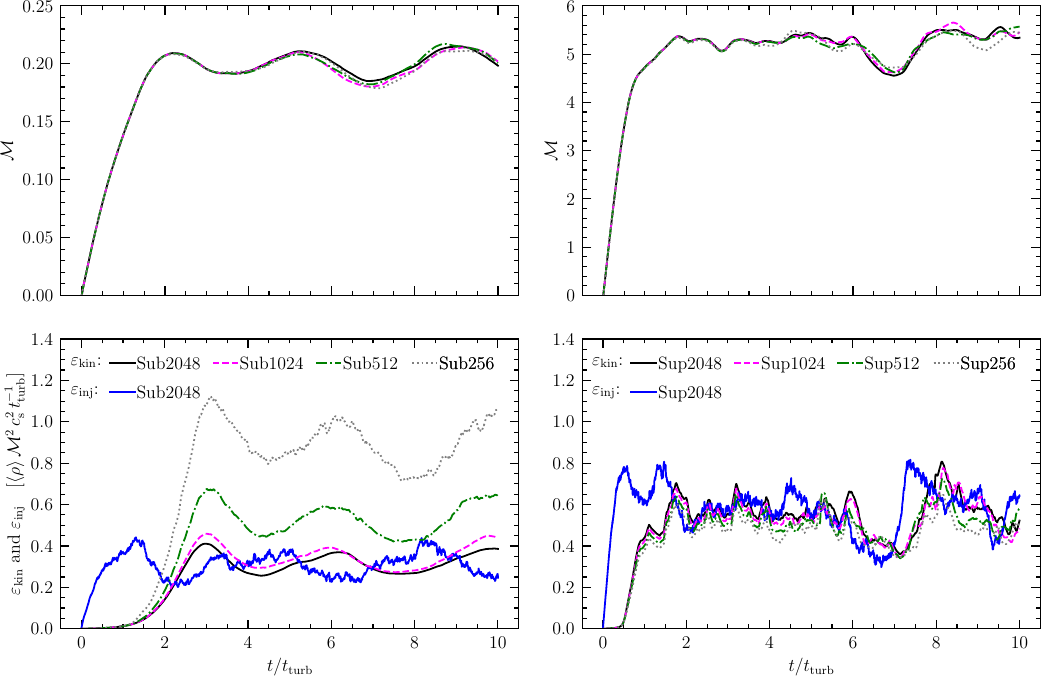} 
\caption{Time evolution of $\mach$ (top panels) and $\epskin$ (bottom panels) for the subsonic regime (left-hand panels) and supersonic regime (right-hand panels) for the $256^3$, $512^3$, $1024^3$, and $2048^3$ simulation resolutions (see legend). Additionally, in the bottom panels, the injection rate $\epsinj$, is shown as a solid blue line for the $2048^3$ runs, noting that $\epsinj$ is prescribed by the turbulence driving (cf.~Sec.~\ref{sec:turbdriv}), and therefore identical across simulations with different resolution. The comparison of $\epsinj$ with $\epskin$ together with the resolution dependence of $\epskin$ emphasises the need for sufficiently high resolution (here $\gtrsim1024^3$) to converge to the physical dissipation rate set by the viscosity $\nu$ and associated target Reynolds number $\Reynolds=2500$. We also note an apparent time delay between the injection and dissipation rate, which is quantified in Fig.~\ref{fig:time_correlation}.}
\label{fig:time evolutions}
\end{figure*}

We begin our quantitative analysis of the simulation suite by investigating the time evolution of the sonic Mach number $\mach$, and the dissipation rate $\epskin$, over the runtime of 10~turnover times, $10\,\tturb$. This is shown in Figure~\ref{fig:time evolutions}. We first examine how resolution affects these quantities and then determine the time correlation between $\epskin$ and the energy injection rate $\epsinj$.

\subsubsection{Time evolution and resolution study of \texorpdfstring{$\mach$}{} and \texorpdfstring{$\epskin$}{}}

The first panel of Figure~(\ref{fig:time evolutions}) shows an initial transient phase, during which the flow is accelerated from rest to reach a fully-developed turbulent state \citep[e.g.,][]{FederrathDuvalKlessenSchmidtMacLow2010,PriceFederrath2010}. This lasts for $\sim2\,\tturb$ in which the target velocity reaches a steady state of $\mach=0.20\pm0.01$ and $\mach=5.2\pm0.3$ for the subsonic and supersonic cases, respectively. We must, therefore, remove this transience from our analysis so that we can focus on the regime of fully-developed turbulence. As a consequence, the rest of this paper will only present analyses based on the data from $2\,\tturb$ onwards.

The rate at which large-scale kinetic energy is injected into the system, $\epsinj$, along with the rate at which this energy dissipates on the small scales, $\epskin$, is shown in the bottom panels of Figure~(\ref{fig:time evolutions}). In the subsonic regime, we observe a large decrease in the magnitude of the dissipation rate as we increase the simulation resolution. Indeed, we find convergence of $\epskin$ with increasing resolution (see measurements of $\epskin$ in column~5 of Tab.~\ref{tab:sim_params}). In particular, $\epskin$ for the simulations with $1024^3$ and $2048^3$ agrees to within $12\%$ compared to $\epskin$ in the $256^3$ to $512^3$ runs, which is different by $40\%$. Any further increases in resolution are limited by computing power and due to this convergence, would be of minimal benefit, at least for the volume-averaged quantities shown here.
 
The resolution dependence suggests that for the lower-resolution runs ($256^3$ and $512^3$), numerical dissipation dominates over the controlled physical dissipation prescribed by $\nu$ (cf.~Sec.~\ref{sec:hd}), and the associated Reynolds number is effectively smaller than our target $\Reynolds=2500$ \citep{ShivakumarFederrath2025}. The consequence of under-resolving the Reynolds number is the nonphysical extra loss of energy seen in the strong mismatch of injection and dissipation rate levels at low resolution. Therefore, we require at least $1024^3$~cells to obtain a reasonably converged dissipation rate at the target $\Reynolds=2500$ in the subsonic regime. Further, by increasing the resolution we introduce more grid cells to average over and significantly smoothen the dissipation rate curve. The injection rate shows a low-frequency variation (of order $1\,\tturb$) and a high-frequency component{\emdash}the former reflects the natural variation of the turbulence modulated via the Ornstein-Uhlenbeck process on $1\,\tturb$, while the latter is due to the {\sc TurbGen} driving pattern only updating $100$~times per $\tturb$ (cf.~Sec.~\ref{sec:turbdriv}).
 
In contrast to the subsonic case, a resolution dependence is not seen in the supersonic simulations, at least not for the resolutions tested here, as evidenced by the bottom, right-hand panel in Figure~\ref{fig:time evolutions}. The main reason for this is that the dissipation scale is resolved and converged faster in the supersonic case, as we will show below.


\subsubsection{Time correlation between injection and dissipation}

We see in the bottom panels of Figure~\ref{fig:time evolutions} that there is a time lag between $\epsinj$ and $\epskin$, in that dissipation seems to lag injection of energy. Large-scale kinetic energy injected into the system falls down the turbulent cascade in some finite time, which we define as the \inquotes{time lag}, $\injdissdt$. We expect some correlation between peaks of dissipation and peaks of injection, which is due to the large-scale driving, $\mathbf{F}$ in Eq.~(\ref{eq:momentum}). This is indeed what is observed in the bottom panels of Figure~(\ref{fig:time evolutions}) throughout the $10\,\tturb$. We determine the time lag between injection and dissipation by first defining the time-lagged difference between injection and dissipation, as
\begin{equation}
q(t,\injdissdt)\equiv\epsinj(t)-\epskin(t+\injdissdt)\label{eq:q_def}.
\end{equation}
We then compute a normalised time-integral of this,
\begin{equation}
    \injdissC(\injdissdt)=\frac{\int\left( q(t,\injdissdt)-\frac{\int q(t,\injdissdt)\,dt}{\int dt}\right)^2\,dt}{\int\epsinj(t)\,dt}\label{eq:C_injdiss},
\end{equation}
which provides the time lag of maximum correlation between injection and dissipation when $\injdissC(\injdissdt)$ is minimal.

In practice, we take the dissipation and injection rate of the Sub/SupN2048 simulations and partition them into two time windows of $1<\tturb<5$ and $5<\tturb<9$ and then shift them by $\injdissdt$, which we continuously increase from $0\,\tturb$ to $2\,\tturb$ in steps of $\tturb/100$. For each $\injdissdt$, we compute $\injdissC$, as shown in Figure~\ref{fig:time_correlation}. From this we find the $\injdissdt$ for which $\injdissC$ is minimised, which yields $\injdissdt/\tturb=1.64\pm0.21$ and $0.48\pm0.07$ for SubN2048 and SupN2048, respectively. We also compute $\injdissdt$ for all other resolutions and report them in column~6 of Tab.~\ref{tab:sim_params}, where we find convergence for $N\gtrsim1024$. As we partition our time domain into two windows, we also obtain an uncertainty in $\injdissdt$ across the two subsets.

Physically, the value of $\injdissdt$ that minimises $\injdissC$ is the time it takes for large-scale injection of energy to cross the turbulent \inquotes{cascade} to where this energy dissipates, which may be on different scales, but likely primarily on small scales. The immediate implication of the measured values of $\injdissdt$ is that dissipation occurs on timescales of the order of $\tturb$, but is a factor of $\sim1.5\times$ slower than $\tturb$ for subsonic turbulence and a factor $\sim2\times$ faster than $\tturb$ for supersonic turbulence. There are three phenomena that may contribute to this. First, it is likely related to the steeper velocity spectrum in supersonic compared to subsonic turbulence \citep{FederrathEtAl2021}. Second, it is possibly a consequence of shocks \inquotes{jumping} the cascade, i.e., introducing non-locality in wave number space, compared to subsonic turbulence, where a $k$-by-$k$ energy transfer is usually regarded as the main driver of the cascade \citep{Frisch1995}. Third, the bottleneck effect \citep{Falkovich1994}, which is stronger in subsonic turbulence, introduces an additional delay before the energy can reach the ultimate dissipation scale.

\subsection{Correlation of dissipation rate with density and vorticity}\label{sec:dens_vort_correlation}

\begin{figure*}
\centering
	\includegraphics[width=1\textwidth]{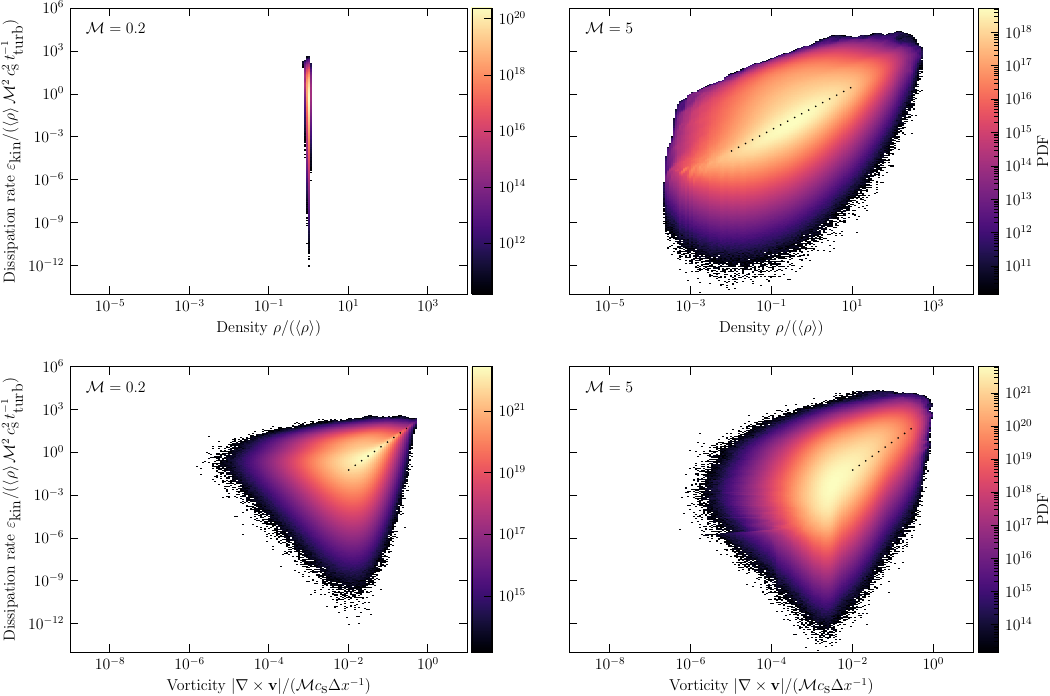} 
\caption{Correlation probability distribution functions between the dissipation rate $\epskin$, and density $\rho$ (top panels) or vorticity $\nabla\times\mathbf{v}$ (bottom panels) in subsonic ($\mach=0.2$, left) and supersonic ($\mach=5$, right) turbulence. The dotted lines show approximate power-law trends. In the supersonic regime, $\epskin$ correlates with density as $\epskin\propto\rho^{3/2}$, consistent with shock-dominated dissipation. In the subsonic regime, dissipation correlates with vorticity as $\epskin\propto|\nabla\times\mathbf{v}|^2$, consistent with small-scale vortex dissipation. The same fit is superimposed in the bottom right panel to facilitate direct comparison between the subsonic and supersonic regime, indicating that some dissipation{\emdash}in addition to shocks{\emdash}stems from vorticity also in the supersonic regime.}
\label{fig:averaged 2d pdfs}
\end{figure*}

Our qualitative analysis based on Figure~\ref{fig:2x2movieframes} suggested a clear correlation of $\epskin$ with $\rho$ in the supersonic case. The subsonic case, however, does not have significant density variations, and based on previous research we rather expect dissipation to be related to vorticity. To provide quantitative measurements of these correlations, we examine 2D (correlation) probability distribution functions of $\epskin$ against density and vorticity, which are shown in Figure~\ref{fig:averaged 2d pdfs}.

Additionally, in Appendix~\ref{sec:diss_correlation} we compute the strain rate dissipation, $\strain$, and show its correlation with $\epskin$.

\subsubsection{Correlation of \texorpdfstring{$\epskin$}{} with \texorpdfstring{$\rho$}{}}
\label{sec:rho2Dpdf}

In the subsonic regime, the dissipation rate shows no correlation with density, as evidenced by the effective $\delta$-function distribution (top left panel). This is consistent with the minimal density variations seen in Figure~\ref{fig:2x2movieframes}, and highlights that density plays no significant role for dissipation in subsonic turbulence.

In contrast, the dissipation rate in the supersonic case has a clear correlation with $\rho$ (top right panel), which we can empirically determine through a power-law relation. By fitting such a line along the ridge of high probability, we find $\epskin\propto\rho^{3/2}$ provides a good representation (see dotted line). Looking back at Fig.~\ref{fig:2x2movieframes}, we see that thin, dense planes of gas (that appear as line filaments in the slices) are strong indicators of a high dissipation, and this correlation is made quantitative through the presented power-law relation.

\subsubsection{Correlation of \texorpdfstring{$\epskin$}{} with \texorpdfstring{$\vort$}{}}

Having established a strong correlation between dissipation and density in the supersonic regime, we now turn to the subsonic case. Here, the relevant quantity is the vorticity of the velocity field. The 2D PDFs of $\epskin$ versus $\vort$ (Figure~\ref{fig:averaged 2d pdfs}, bottom-left) reveal a clear correlation, which is well described by a power law $\epskin\propto\vort^2$. This scaling is consistent with dissipation arising from vortex dissipation and interactions along the turbulent cascade. The same fit is shown in the supersonic case (bottom-right) for comparison, although the absence of a clear ridge indicates that no unique power-law relation can be defined there.

The $\epskin\propto\vort^2$ dependence implies that regions of strongest vorticity coincide with the sites of highest dissipation in subsonic turbulence \citep{GotohWatanabeSaito2023}. Energy transfer through the cascade is therefore regulated by vortex interactions on progressively smaller scales, occurring on relatively long timescales. This provides a natural explanation for the factor of $\sim3$ longer dissipation time lag in the subsonic regime compared to the supersonic case (see Section~\ref{sec:timevol}). In supersonic turbulence, collisions of dense sheets generate localised regions of high vorticity, but the dominant role of shocks accelerates energy transfer and shortens the dissipation timescale.

\subsection{Velocity and dissipation rate spectra}\label{sec:spectra}

\begin{figure*}
\centering
	\includegraphics[width=1\textwidth]{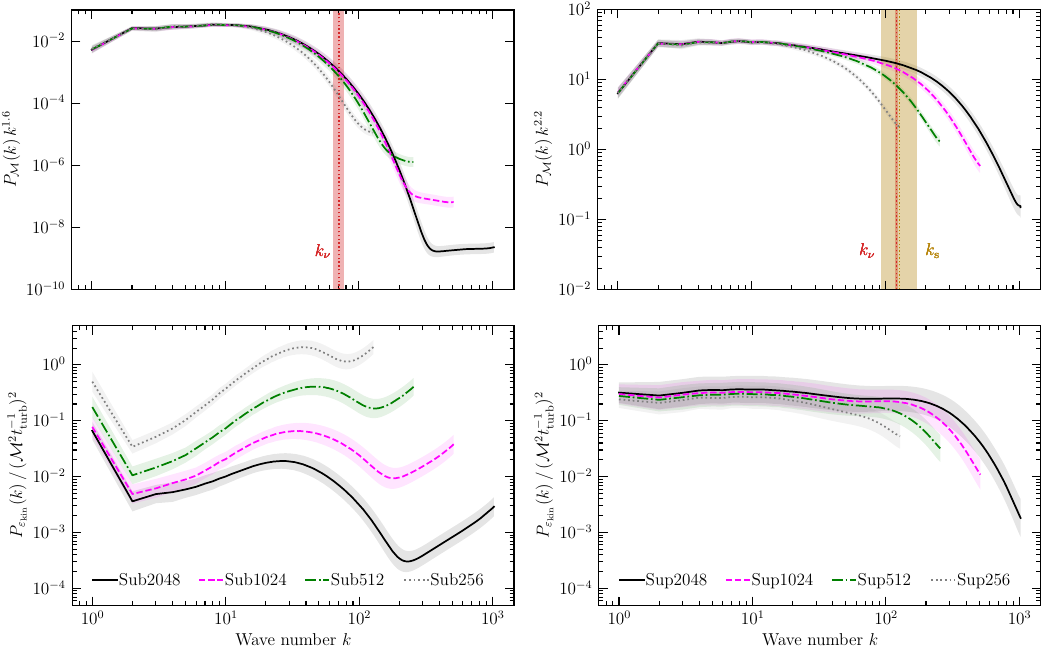} 
\caption{Velocity ($\mathcal{M}$; top panels) and dissipation rate ($\epskin$; bottom panels) power spectra for subsonic (left panels) and supersonic (right panels) turbulence. The velocity spectra (top panels) are compensated by $k^{-1.6}$ and $k^{-2.2}$, respectively, to emphasise the approximate self-similar scaling range (\inquotes{inertial range}). The driving scale at $k=2$ is visible in all spectra, while the viscous dissipation wave number $k_\nu$ \citep{KrielEtAl2025} is marked as a red vertical band. In the supersonic case the sonic scale $k_\mathrm{s}$ \citep{FederrathEtAl2021} is indicated as a gold vertical band. The resolution dependence shows that we require $N\gtrsim1024$ for the velocity spectra to converge on scales $k\lesssim k_\nu$ in both the subsonic and supersonic regime. The dissipation spectra (bottom panels) are normalised by $(\propto\mach^2\tturb^{-1})^2$ to enable direct comparison between regimes. Subsonic dissipation fails to converge even at $2048^3$. However, the peak dissipation{\emdash}while not fully converged either{\emdash}is relatively stable around \mbox{$k\sim30$--$50$}, indicative of dissipation concentrated on small scales, i.e., relatively large wave numbers close to, but somewhat smaller than $k_\nu$. In contrast, the supersonic spectra converge on scales $k\lesssim k_\nu$ at $\sim1024^3$, with a relatively flat dissipation spectrum. There is some indication of two weak local peaks near $k\sim10$ and $k\sim k_\nu$, which may reflect the combination of elongated shocks extending through large fractions of the domain, and their thin widths. In addition, the supersonic case also has contributions from vortex dissipation on small scales. This demonstrates the fundamentally different dissipation behaviour: subsonic turbulence is dominated by small-scale dissipation, while supersonic turbulence exhibits dissipation across a broader range of scales.}
\label{fig:power spectra}
\end{figure*}

We now turn our attention to the power spectra of $\mach$ and $\epskin$. This provides insight into the scale at which these quantities are strongest. For the velocity field, we expect inertial-range slopes consistent with $k^{-5/3}$ in the subsonic regime \citep{Kolmogorov1941c}, although subject to intermittency corrections \citep{SheLeveque1994} and $k^{-2}$ in the supersonic, shock-dominated case \citep{Burgers1948,Federrath2013,FederrathEtAl2021}. We further examine the dissipation rate spectra, motivated by the expectation that dissipation in subsonic turbulence is concentrated on small scales, scaling as $\epskin\propto\vort^2$, whereas in supersonic turbulence it may extend across a broader range of scales due to the density dependence $\epskin\propto\rho^{3/2}$. For both of these spectra, we provide a resolution study across our simulation suite.

\subsubsection{Power spectra of \texorpdfstring{$\mach$}{}}

The velocity spectra are shown in the top panels of Figure~\ref{fig:power spectra}, compensated by $k^{-1.6}$ and $k^{-2.2}$ in the subsonic and supersonic regime, respectively. The chosen compensations are close to the expected power-law scaling in the respective regimes{\emdash}these are $\sim k^{-5/3}$ for subsonic turbulence \citep{Kolmogorov1941c,SheLeveque1994}, and $\sim k^{-2}$ for supersonic, shock-dominated turbulence \citep{KritsukEtAl2007,PriceFederrath2010,Federrath2013}. As a result of our fixed Reynolds number, $\Reynolds=2500$, we expect the dissipation wave number to be $k_\nu=\kdriv\Reynolds^{3/4}=77.8\pm 7.1$ and $k_\nu=\kdriv\Reynolds^{2/3}=121.6\pm3.7$, in the subsonic and supersonic case, respectively \citep{KrielEtAl2025}. These are shown as vertical dotted lines with a shaded region representing the 16th to 84th percentile range around the mean $k_\nu$ value. The resolution study shows only the simulations with $N\gtrsim1024$ are sufficiently converged on scales $k\lesssim k_\nu$, in terms of the velocity spectra. For the supersonic case, we also show the sonic scale $k_\mathrm{s}$

In the subsonic case, it is clear that $k_\nu$ is overestimated which is likely due to a low Reynolds number. However, the supersonic case demonstrates the necessity of a $2048^3$ resolution simulation as we see that for each increase in resolution, the spectra converge such that the dissipation wave number is properly resolved.

\subsubsection{Power spectra of \texorpdfstring{$\epskin$}{}}\label{sec:spectra_epskin}

The bottom panels of Figure~\ref{fig:power spectra} present the power spectra of the dissipation rate. We do not compensate these spectra but instead normalise them by units of $\epskin^2$ so that the subsonic and supersonic dissipation rates are comparable. It is clear that the subsonic spectra do not converge, even for $N=2048$. It should also be noted that the resurgence in the spectrum at grid size scales $(k\sim 10^3)$ is a result of numerical effects. We see these numerical artefacts as very fine structures of size $\sim1\dx$ (cf.~Fig.~\ref{fig:movie_frame_res}). This highlights that the dissipation in this regime is hard to capture even for a moderately low Reynolds number of $2500$ and a relatively large grid resolution. We only see convergence on the box scale $(k=1)$, i.e., at least the overall dissipation rate is reasonably converged, consistent with what we found for the time evolution in Figure~\ref{fig:time evolutions}. Thus, we conclude that extremely high resolution is required to fully converge on the dissipation properties of subsonic turbulence, motivating future investigation. Despite the strong resolution dependence in this regime, we can still draw some broad conclusions on dissipation in subsonic turbulence: it is primarily concentrated on small scales, indicated by the peak in $\epskin$ at \mbox{$k\sim30$--$50$}, somewhat less than the dissipation wave number estimated in \citet{KrielEtAl2025} given our Reynolds number.

In contrast, the $\epskin$ spectra for supersonic turbulence converge for $N\gtrsim1024$. We find that dissipation occurs relatively evenly throughout the entire scale range. However, we see some indication of two weak local maxima around $k\sim10$ and $100$. Given the dominance of shocks in this regime, we expect density structures that are both thin (small width) and long. Thus, given the correlation, $\epskin\propto\rho^{3/2}$ that we determined in Section~\ref{sec:rho2Dpdf} we could associate these two peaks with the length and width of the density structures (shocks) that cause most of the dissipation in the supersonic regime. However, a rigorous verification according to \citet{Roy2019} is left for future work. In addition, there is some small-scale contribution from vortex dissipation around $k_\nu\sim k_\mathrm{s}$, which may also contribute to the mild peak around $k\sim100$.

In summary, we conclude that dissipation in supersonic turbulence is largely scale independent, whereas dissipation in subsonic turbulence is predominately a small-scale quantity with a distinct peak at $k\lesssim k_\nu$.

\subsection{Fractal dimension analysis}\label{sec:fracdim}

\begin{figure*}
\centering
	\includegraphics[width=1\textwidth]{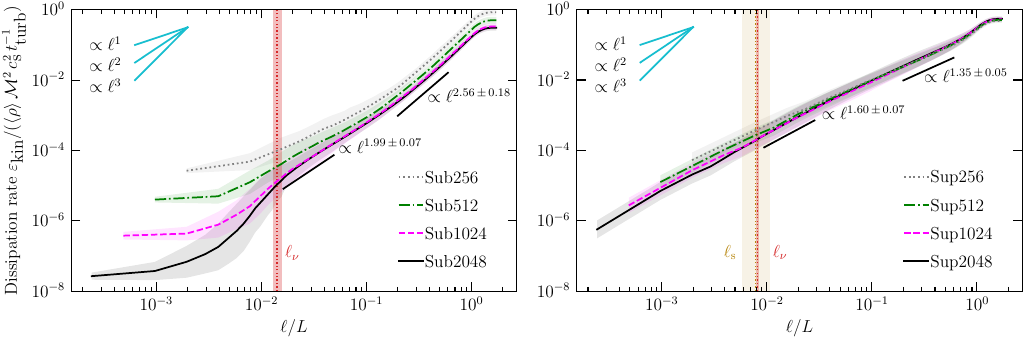} 
\caption{Fractal dimension analysis of the dissipation rate for subsonic (left) and supersonic (right) turbulence and for different grid resolutions. The shaded regions indicate $1\sigma$ temporal fluctuations. Vertical lines denote the viscous dissipation scale, $\ell_\nu$ (red), and for the supersonic case, the sonic scale, $\ell_\mathrm{s}$ (gold). Power-law fits are shown with slopes annotated in each panel. In the subsonic regime, the measured fractal dimension, $D$, varies strongly with scale. Around $\ell_\nu$, the slope corresponds to $D=1.99\pm0.07$, indicative of sheet-like vortex dissipation structures, while at larger fractions of the box ($\ell/L\in[0.2,0.6]$) we obtain $D=2.56\pm0.18$, consistent with more volume-filling dissipation. In the supersonic regime, the fractal dimension is nearly scale-invariant. At $\ell_\nu$ we measure $D=1.60\pm0.07$, suggesting a mixture of sheet- and filament-like structures, with sheets somewhat dominating. On larger scales the fractal dimension decreases to $D=1.35\pm0.05$, indicating that dissipation becomes increasingly filamentary. This behaviour is consistent with dissipation concentrated at the intersections of strong, long shocks. Resolution dependence is evident in both regimes, but stronger for the subsonic case: convergence requires $N\gtrsim1024^3$ to converge on scales $\ell\gtrsim\ell_\nu$ (cf.~Tab~\ref{tab:sim_params}).}
\label{fig:fracdim}
\end{figure*}

Our qualitative analysis in Figure~\ref{fig:2x2movieframes} suggested that for $\mach=0.2$ the dissipation is relatively volume filling on large scales and may be associated with vortex tubes (filaments) and shearing sheets on small scales \citep{GotohWatanabeSaito2023}. The $\mach=5$ case on the other hand is dominated by shocks. In isolation, these shocks correspond to 2D planes, however, when they collide they form 1D filaments. Therefore, to quantify the dimensionality of these dissipation structures on different scales, we now perform a fractal dimension analysis of $\epskin$ in both the subsonic and supersonic regime.

\subsubsection{Method}

Multiple definitions and estimators have been proposed to quantify the fractal dimension of hierarchical structures such as turbulence, including the box-counting, perimeter–area, and mass–radius methods \citep{Mandelbrot1975,MandelbrotWheeler1983,Meyers1992,SanchezEtAl2005,KritsukEtAl2007,FederrathKlessenSchmidt2009}.
Given that $\epskin$ is a real-valued field embedded in 3D, we adopt the mass–radius approach, constructing $\epskin(\ell)$ and inferring the scaling exponent from the relation
\begin{equation} \label{eq:fracdim}
\epskin(\ell) \propto \ell^D,
\end{equation}
where $\ell$ denotes the diameter of a spherical region and $D$ is the inferred fractal dimension. In general, $D\to D(\ell)$ may itself vary with $\ell$, in which case the field is referred to as multifractal. It has been shown that multifractal models of turbulence can also be theoretically linked to the Navier-Stokes equations \citep{DubrulleGibbon2022}.

We begin by identifying the grid cell with the maximum value of $\epskin$. Concentric spheres of increasing diameter $\ell$ are then constructed around this position, with $\ell$ varied to span the entire domain, i.e.\ $\ell \in [\dx, L]$ in increments of $2\dx$. For each sphere, we compute and record the total enclosed $\epskin$, applying triply periodic boundary conditions to avoid truncation at the domain boundaries. A linear regression is then performed in log–log space over a local range of $\ell$ to determine $D$ within that interval. Specifically, we measure $D$ near the dissipation scale $\ell_\nu$ \citep[estimated for $\Reynolds=2500$ following][]{KrielEtAl2025} and on large scales, $0.2\leq\ell/L\leq0.6$, to quantify the global, volume-filling dimensionality of $\epskin$.

We note for context that an alternative, complementary route to probing the geometry of dissipation is via intermittency analysis of velocity structure functions \citep{Frisch1995}. In multifractal models, the anomalous scaling exponents of the velocity structure functions can be related to the spectrum of singularities of the dissipation field, and hence to its effective fractal (or multifractal) dimension. This approach has been used extensively in subsonic turbulence \citep{SheLeveque1994,DubrulleGibbon2022} and extended to supersonic regimes relevant for molecular clouds \citep{Boldyrev2001,Schmidt2008,KonstandinEtAl2012}. A systematic intermittency-based comparison of our datasets is left for future work.

\subsubsection{Results}\label{sec:fracdim_results}

Figure~\ref{fig:fracdim} shows $\epskin(\ell/L)$ for the subsonic and supersonic case, and for different numerical resolutions. Broadly comparing the subsonic and supersonic cases, we see that there is considerably more resolution- and scale-dependence of the fractal dimension $D$ than in the supersonic case, which shows a relatively uniform power-law scaling. We quantify $D$ by fitting power laws with Eq.~(\ref{eq:fracdim}) in two regions of $\ell$: one on small scales, near the dissipation scale, $\ell_\nu$, where dissipation is maximal (cf.~Fig.~\ref{fig:power spectra}), and another one on large scales to quantify the overall volume-filling dimensionality of the dissipation structures. Both power-law fits are performed over a factor of $3$ in scale, providing sufficient scale range to fit a power law data and at the same time providing a scale-local measure of $D$.

In the case of subsonic turbulence we find $D=1.99\pm0.07$ near $\ell_\nu$, i.e.~close to 2D, which implies that intense dissipation is predominantly organised in thin shear layers that appear as ribbon- or sheet-like structures. These sheets occur in close association with flattened vortex tubes, consistent with experimental and numerical studies showing clusters of vortex tubes embedded in layer-like regions of strong shear, with sheet-like dissipation adjacent to the tubes \citep{MoffattKidaOhkitani1994,
MoisyJimenez2004,FiscalettiWesterweelElsinga2014}. In Figure~\ref{fig:2x2movieframes} (top left panel), we saw that the dissipation is volume filling on scales that are comparable to the box size. Thus, another power law is fitted for $0.2\leq\ell/L\leq0.6$, which indicates a fractal dimension of $D=2.56\pm0.18$ in the subsonic case, indicative of a fairly volume-filling quantity. While we do not perform an explicit fit for the fractal dimension for $\ell\lesssim\ell_\nu$ due to non-convergence for $\ell\lesssim\ell_\nu$, we see an indication that on very small scales, the data is compatible with $D\sim1$, which indicates thin vortex filaments{\emdash}however, this result is not conclusive given the large resolution dependence in that scale range.

Conversely, the fractal dimension of the supersonic dissipation rate is relatively invariant throughout all scales. Indeed, we see that near the dissipation scale $\ell_\nu$, the fractal dimension is $D=1.60\pm 0.07$, which suggests that the dissipation rate is neither exactly 1D nor 2D, but instead a combination of both. This indicates that the structures formed are a mixture of dissipation sheets ($D=2$) \citep{RichardLesaffreFalgaroneLehmann2022} and filaments ($D=1$), both due to shocks surfaces, where the latter represents intersections of those. As this balance is not even (i.e., $D\gtrsim1.5$), suggesting that sheets are slightly more frequent than filaments. This is consistent with the fact that an intersection requires at least two dissipation sheets. For the large-scale power law fit in the region of $[0.2,0.6]\ell/L$ we find $D=1.35\pm0.05$, lower than in the dissipation range. This indicates that on box size scales, the dissipation is not predominately structured in planes, but is now more weighted towards filamentary structures. Finally we note that the supersonic case may also have contributions from vortex dissipation on small scales (cf.~bottom right-hand panel of Fig.~\ref{fig:averaged 2d pdfs}), and hence, $D$ may be somewhat biased towards more 2D structures on small scales, similar to the subsonic case. This contribution from the vorticity arises due to a logarithmic density gradient term in the vorticity evolution equation \citep{MeeBrandenburg2006,Federrath2011} meaning regions of high density variation (e.g.~shocks) lead to the observed vorticity. This is further supported by the sonic scale $\ell_\mathrm{s}$ shown in Fig.~\ref{fig:fracdim}, which is in this case located close to $\ell_\nu$.

\section{Conclusions} \label{sec:conclusions}

In this work, we used high-resolution numerical simulations to study the statistics and structure of dissipation in subsonic and supersonic turbulence. We solve the hydrodynamic equations with explicit viscosity corresponding to a Reynolds number of $\Reynolds=2500$ for different grid resolutions from $256^3$ to $2048^3$ cells to study numerical convergence of all analysis quantities, in particular the kinetic energy dissipation rate, $\epskin$. We quantify its morphology (Section~\ref{sec:movie}), time evolution (Section~\ref{sec:timevol}), correlation with density and vorticity (Section~\ref{sec:dens_vort_correlation}), power spectra (Section~\ref{sec:spectra}), and fractal dimension (Section~\ref{sec:fracdim}). Our key results are summarised below:

\begin{enumerate}

    \item We assessed the morphology of the dissipation rate, and the density through visualisations (see Figure~\ref{fig:2x2movieframes}). In the subsonic case, we find that there is qualitatively no correlation between $\epskin$ and $\rho$ but instead with vorticity, $\vort$. This differs from the supersonic case in which we observe a clear correlation between $\epskin$ and $\rho$ due to the presence of shocks.

    \item We then considered the time evolution of the sonic Mach number and dissipation rate. We find that fully-developed turbulence is achieved at $t=2\,\tturb$ and only use data beyond this time for subsequent analyses. The time lag between a kinetic energy injection event, and its eventual dissipation is measured as $\injdissdt/\tturb = 1.64\pm0.21$ for $\mach=0.2$ and $\injdissdt/\tturb = 0.48\pm0.07$ when $\mach=5$. This indicates that given some large-scale injection event, we expect that energy to dissipate a factor $\sim3\times$ faster in supersonic turbulence compared to subsonic turbulence.

    \item A quantitative correlation study between $\epskin$ and both the density $\rho$, and vorticity $\vort$ is conducted through PDFs (see Figure~\ref{fig:averaged 2d pdfs}). We confirm that the subsonic dissipation rate has no correlation with density but instead follows a relationship $\epskin\propto\vort^2$; in particular, it is the rate of strain that correlates with the square of vorticity. Conversely, the dissipation rate in supersonic turbulence primarily correlates with the density as $\epskin\propto\rho^{3/2}$, and there is additional weaker correlation also with the vorticity or rate of strain.

    \item Fourier spectra of the velocity field (see Figure~\ref{fig:power spectra}) are broadly consistent with previous works \citep{Kolmogorov1941c,Burgers1948} but require a compensation of $k^{-1.6}$ and $k^{-2.2}$ to flatten the scaling (\inquotes{inertial}) range in subsonic and supersonic regimes, respectively. The spectra of the dissipation rate reveal that dissipation in subsonic turbulence is remarkably challenging to properly simulate, showing almost no convergence at high wave numbers, even at a $2048^3$ grid resolution and moderate Reynolds number of $2500$, motivating future work. Nonetheless, a peak in the spectra at around \mbox{$k\sim30$--$50$} shows that the dissipation is restricted to small scales in subsonic turbulence. The supersonic case instead shows that dissipation is present at effectively the same power on all scales, with two weak peaks at around $k\sim 10$ and $k\sim 10^2$ indicating that shocks causing dissipation $\epskin\propto\rho^{3/2}$ are both thin and long structures with an added small-scale dissipation component from vortices.

    \item Finally, a fractal study (see Figure~\ref{fig:fracdim}) of the dissipation rate is used to determine the fractal dimension of $\epskin$ near the dissipation scale $\ell_\nu$, and also on box-size scales. In subsonic turbulence we find $D\sim2$ near $\ell_\nu$, indicating that intense dissipation is predominantly organised in thin shear layers associated with flattened vortex tubes, consistent with previous experimental work \citep{FiscalettiWesterweelElsinga2014}. In contrast, the dissipation rate in supersonic turbulence has a dimensionality of $D\sim1.6$ near the dissipation scale and $D\sim1.4$ on large scales, reflecting that dissipation occurs primarily in shock sheets ($D=2$) and their filamentary intersections ($D=1$), consistent with the visual analysis of $\epskin$ in Figure~\ref{fig:2x2movieframes}.
\end{enumerate}

The results of this work quantify and emphasise the fundamentally distinct statistics and structure of kinetic energy dissipation in subsonic and supersonic turbulence. This may provide theoretical input for studies on the heating properties of turbulence in the interstellar medium, its locality, scale dependence, structure, and statistics. Future work needs to address the challenges in achieving numerical convergence of the wave number-dependence of the dissipation rate in subsonic turbulence, as well as the role of magnetic fields, i.e., by solving the full magnetohydrodynamic equations. As an extension of the analysis we provide, one could quantify dissipation localisation further by computing the fraction of total dissipation rate, $\epskin$, arising from the most dissipative volume fraction, following \citet{RichardLesaffreFalgaroneLehmann2022}. This is particularly relevant for modelling chemistry/phase changes driven by intermittent heating.

\section*{Acknowledgements}
E.~T.~acknowledges funding via an ANU RSAA Science PhB Advanced Studies Course scholarship. C.~F.~acknowledges funding provided by the Australian Research Council (Discovery Project grants~DP230102280 and~DP250101526), and the Australia-Germany Joint Research Cooperation Scheme (UA-DAAD). We further acknowledge high-performance computing resources provided by the Leibniz Rechenzentrum and the Gauss Centre for Supercomputing (grants~pr32lo, pr48pi, pn76ga and GCS Large-scale project~10391), the Australian National Computational Infrastructure (grant~ek9) and the Pawsey Supercomputing Centre (project~pawsey0810) in the framework of the National Computational Merit Allocation Scheme and the ANU Merit Allocation Scheme. The simulation software, \texttt{FLASH}, was in part developed by the Flash Centre for Computational Science at the University of Chicago and the Department of Physics and Astronomy at the University of Rochester.

\section*{Data Availability}
The simulation data ($\sim\!100\,\mathrm{TB}$) will be shared on reasonable request to the authors.


\newcommand{\rmp}{Rev.~Mod.~Phys.}

\bibliographystyle{mnras}
\bibliography{troccoli} 




\appendix

\section{Derivation of kinetic energy equation} \label{sec:derivation}

Here we provide the derivation of the kinetic energy equation in Section~\ref{sec:ekin}. To proceed, we will require both Eqs.~\eqref{eq:masscons} and \eqref{eq:momentum}. The kinetic energy is $\propto\rho\abs{\mathbf{v}}^2$, so first take the dot product of Eq.~\eqref{eq:momentum} with $\mathbf{v}$ and observing that it can also be conveniently expressed in component form as
\begin{equation}
v_i \partial_t(\rho v_i) + v_i \partial_j (\rho v_i v_j) = - v_i \partial_i p.
\end{equation}
Applying the product rule and using $v_i \partial_t v_i = \tfrac{1}{2}\partial_t v_i^2$ and $v_i \partial_j v_i = \tfrac{1}{2}\partial_j v_i^2$, we obtain
\begin{equation}
\frac{1}{2}\rho(\partial_t v_i^2 + v_j \partial_j v_i^2) + \frac{1}{2}v_i^2(\partial_t \rho + \partial_j (\rho v_j)) = -v_i \partial_i p.
\end{equation}
The continuity equation eliminates the second term. Recognising the remaining terms as a divergence, we write
\begin{equation}
\partial_t\left(\tfrac{1}{2}\rho v^2\right) + \partial_j\left(\tfrac{1}{2}\rho v^2 v_j\right) = - v_i \partial_i p.
\end{equation}
Defining the kinetic energy density $\ekin = \tfrac{1}{2}\rho v^2$, this becomes
\begin{equation}
\partial_t \ekin + \nabla \cdot (\ekin \mathbf{v}) = -\mathbf{v}\cdot\nabla p.
\end{equation}
Finally, using $\nabla \cdot (p \mathbf{v}) = p\nabla\cdot \mathbf{v} + \mathbf{v}\cdot\nabla p$, we obtain
\begin{equation}
\partial_t \ekin + \nabla \cdot \left[(\ekin + p)\mathbf{v}\right] = p \nabla \cdot \mathbf{v},
\end{equation}
as stated in Section~\ref{sec:epskin}.

\section{Calculation of \texorpdfstring{$\injdissC$}{} for time delay between \texorpdfstring{$\epskin$}{} and \texorpdfstring{$\epsinj$}{}}
\label{sec:l2norm}

\begin{figure*}
\centering
	\includegraphics[width=1\textwidth]{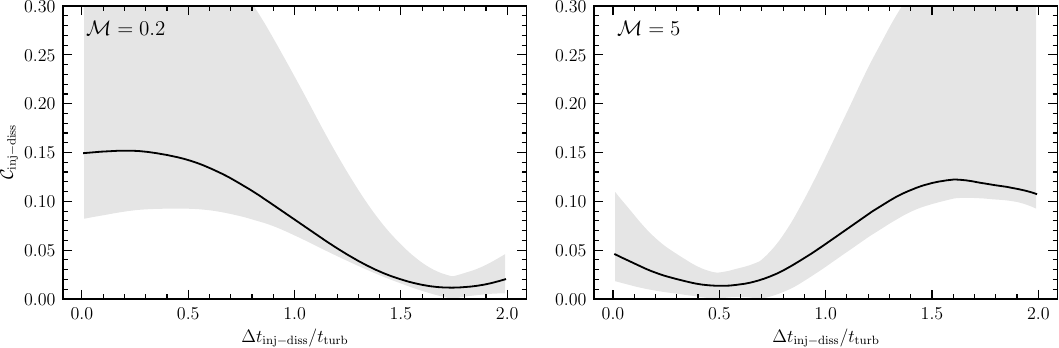} 
\caption{The quantity $\injdissC$ for different time lags. The minimum of $\injdissC$ defines the maximum correlation between $\epskin$ and $\epsinj$ at a time lag of $\injdissdt/\tturb=1.64\pm0.21$ and $0.48\pm0.07$ for the $N=2048$ subsonic and supersonic simulation, respectively.}
\label{fig:time_correlation}
\end{figure*}

To determine the minimal time lag between the injection and dissipation rate, we compute their $\injdissC$ for different time lags. We do this for two separate time windows (see discussion in Section~\ref{sec:timevol}). This gives us two sets of $\injdissC$. Figure~\ref{fig:time_correlation} shows $\injdissC$ (as defined in Eq.~\eqref{eq:C_injdiss}) for the average of these two time windows as a solid line and the shaded region reflects the minimum and maximum of $\injdissC$ in the two time windows. The best value of $\injdissdt$ is determined as the minimum of the solid line, while its uncertainty is found from the minimum of $\injdissC$ between the two time windows.

\section{Resolution dependence of dissipation rate}

\begin{figure*}
\centering
	\includegraphics[width=1\textwidth]{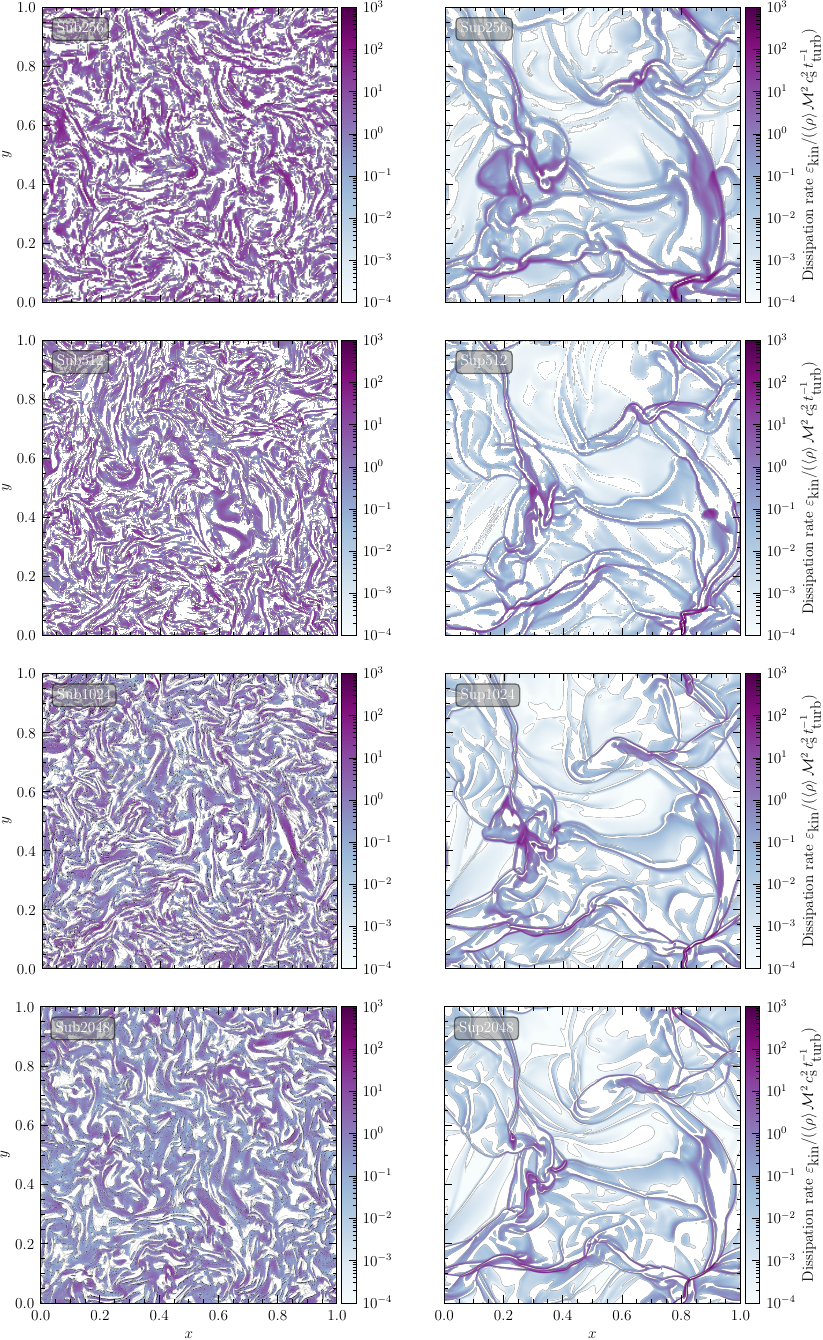} 
\caption{Slices of $\epskin$ through the $xy$ plane in both subsonic (left) and supersonic (right) regimes for all simulation models with different grid resolution (from top to bottom, as indicated in the legends).}
\label{fig:movie_frame_res}
\end{figure*}

Figure~\ref{fig:movie_frame_res} shows the same as Fig.~\ref{fig:2x2movieframes} top panels, but for different numerical resolution $N$ (see Tab.~\ref{tab:sim_params}).

\section{Correlation of \texorpdfstring{$\epskin$}{} and \texorpdfstring{$\strain$}{}}\label{sec:diss_correlation}

\begin{figure*}
\centering
	\includegraphics[width=1\textwidth]{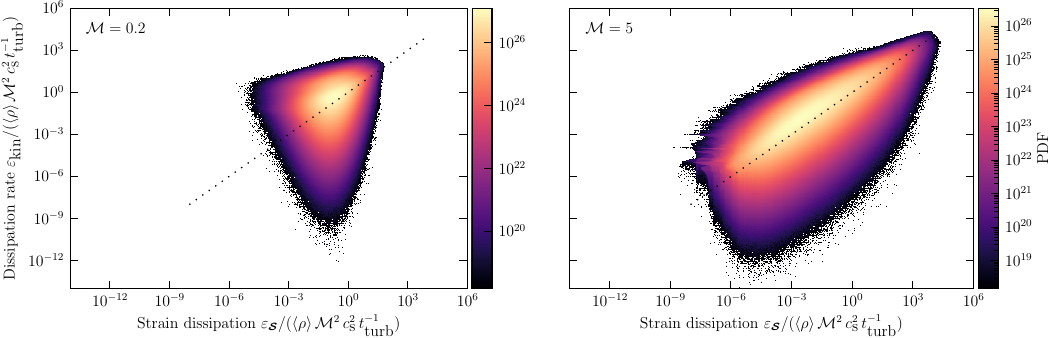} 
\caption{Same as Fig.~\ref{fig:averaged 2d pdfs}, but for the strain rate, $\strain$ (as defined in Eq.~\eqref{eq:strain}). The dotted lines show the one-to-one correlation.}
\label{fig:averaged 2d pdfs strain}
\end{figure*}

The correlation of $\epskin$ with $\vort$ we found in Fig.~\ref{fig:averaged 2d pdfs} arises due to most of the dissipation occurring in shear layers that wrap around vortex tubes where the turbulent rate of strain is strongest \citep{MoffattKidaOhkitani1994}. Therefore, it is interesting to investigate the correlation of $\epskin$ with the rate of strain. Figure~\ref{fig:averaged 2d pdfs strain} presents the correlation between $\epskin$ and the strain rate, 
\begin{equation}
    \strain=2\mu\mathcal{S}_{ij}\mathcal{S}_{ij}\label{eq:strain}.
\end{equation}
In the supersonic regime, there is a clear one-to-one correspondence over the bulk of the distribution, indicating that $\epskin$ tracks strain-based dissipation with relatively small scatter. The subsonic regime presents a weaker, but still significant correlation of $\epskin$ and $\strain$.

\bsp	
\label{lastpage}
\end{document}